\documentclass[3p,times,authoryear,onecolumn]{elsarticle}

\usepackage{subfig}
\usepackage[export]{adjustbox}
\usepackage{color}

\journal{Icarus}
\begin{document}

\begin{frontmatter}

\title{Cohesive Regolith on Fast Rotating Asteroids}

\author[address1]{Paul S{\'a}nchez\corref{correspondingauthor}}
\ead{ddsanche@colorado.edu}
\cortext[correspondingauthor]{Corresponding author}
\address[address1]{Colorado Center for Astrodynamics Research, University of Colorado Boulder, 431 UCB, Boulder, CO 80309, USA}

\author[address2]{Daniel J. Scheeres}
\address[address2]{Aerospace Engineering Department, University of Colorado Boulder, 429 UCB, Boulder, CO 80309, USA}

\begin{abstract}
The migration of cohesive regolith on the surface of an otherwise monolithic or strong asteroid is studied using theoretical and simulation models. The theory and simulations show that under an increasing spin rate (such as due to the YORP effect), the regolith covering  is preferentially lost across certain regions of the body. For regolith with little or no cohesive strength, failure occurs by landsliding from the mid latitudes of the body at high enough spin rates.  As the cohesive strength of the regolith increases, failure occurs by fission of grains (or coherent chunks of grains) across a greater extent of latitudes and eventually will first occur at the equator. As the spin rate is further increased, failure regions migrate from the first failure point to higher and lower latitudes. Eventually failure will encompass the equatorial region, however there always remains a region of high latitudes (around the poles) that will not undergo failure for arbitrarily high spin rates (unless disturbed by some other phenomenon). 
With these results a scaling law is derived that can be used to determine whether observed asteroids could retain surface regolith grains of a given size.  The implications of this for the interpretation of spectral observations of small asteroids and boulder migration on large asteroids are discussed.

%

\end{abstract}


\end{frontmatter}


\section{Introduction}

The current understanding of small asteroids in the Solar System is that they are gravitational aggregates held together by gravitational, cohesive and adhesive forces \citep{science_fujiwara, science_yano, thomas_eros_craters, richardson2009, scheeres2010, sanchez2014, sanchez2016}.  Results from the Hayabusa mission to Itokawa along with in-situ, thermal and radar observations of asteroids have shown that they can be covered in a size distribution of grains that spans from microns to tens of meters \citep{itokawa_boulders,itokawa_miyamoto}.  
Given the regolith-rich surface of these bodies, it is now an open question whether even smaller bodies, down to a few meters in size, could also retain regolith covering. The question is especially compelling for the small-fast and -superfast rotators (SFRs), whose surface centripetal accelerations exceed their gravitational attraction \citep{polishook2013}. 
{\color{black} 
In this paper the mechanics of regolith on a strong subsurface is studied when cohesion is taken into account. It is shown that when cohesion is accounted for, it becomes possible for small-fast rotators to retain regolith up to high spin rates. The current theory provides an extension of the cohesionless theory outlined earlier in \cite{scheeres2015}, and indeed the limiting cases of no cohesion in this current paper generally agree with the earlier study and subsequent simulations \cite{yu2018}. 
}

This paper does not target any specific asteroid, as their shapes and specific characteristics would unnecessarily increase the complexity of the study.  Therefore, it assumes a spherical monolith covered by a thin layer of regolith.  The following section will describe the physical model, which will be followed by the Coulomb theory section that analyzes the conditions for failure from a theoretical point of view.  In the following sections this model is studied through simulations, and thus the numerical method and the simulation setup are described.  Finally,the simulations are compared to the analytical predictions and the implications for SFRs in the NEO population are discussed.

\section{The Model}

The model used for this study is introduced and explained in \cite{scheeres2015}.  
The current paper expands that analysis by adding the effect of cohesion into the failure of surface regolith, and by introducing a simulation methodology for studying the failure of surface regolith on fast spinning bodies using a precise model. 

\begin{figure}[htbp]
\begin{center}
\includegraphics[scale=0.75]{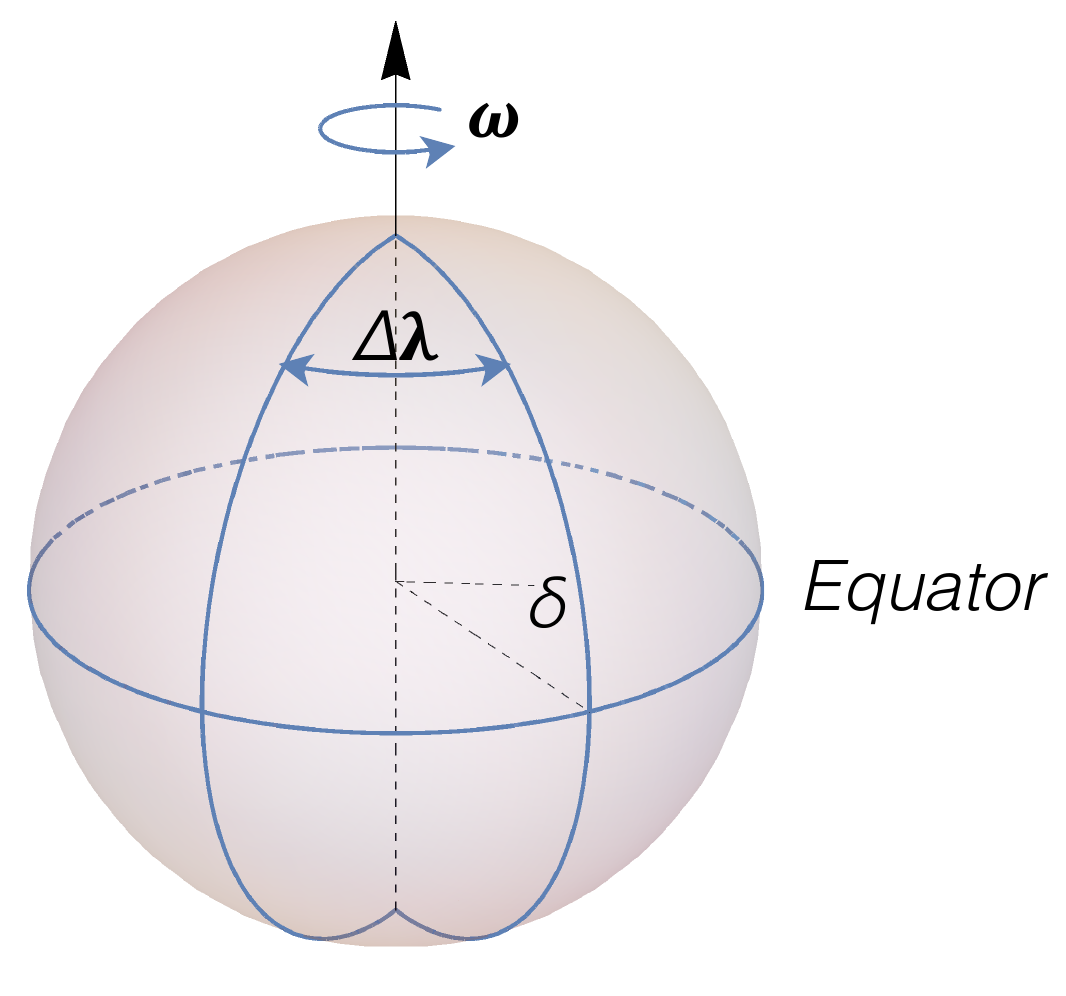}
\caption{Geometry of the test body.}
\label{sphere}
\end{center}
\end{figure}

Figure \ref{sphere} shows the main geometry and variables used in the model.  For simplicity, a sphere is used to represent a regolith or boulder covered asteroid that from now on is called the test body.  This test body is rotating with angular velocity $\bf\omega$ as shown in the figure. 
For the analytical and simulation model consider a lune of this body of angular width $\Delta\lambda$ where the latitude of any point is defined by its latitude $\delta$.

\begin{figure}[h]
\begin{center}
\includegraphics[scale=0.65]{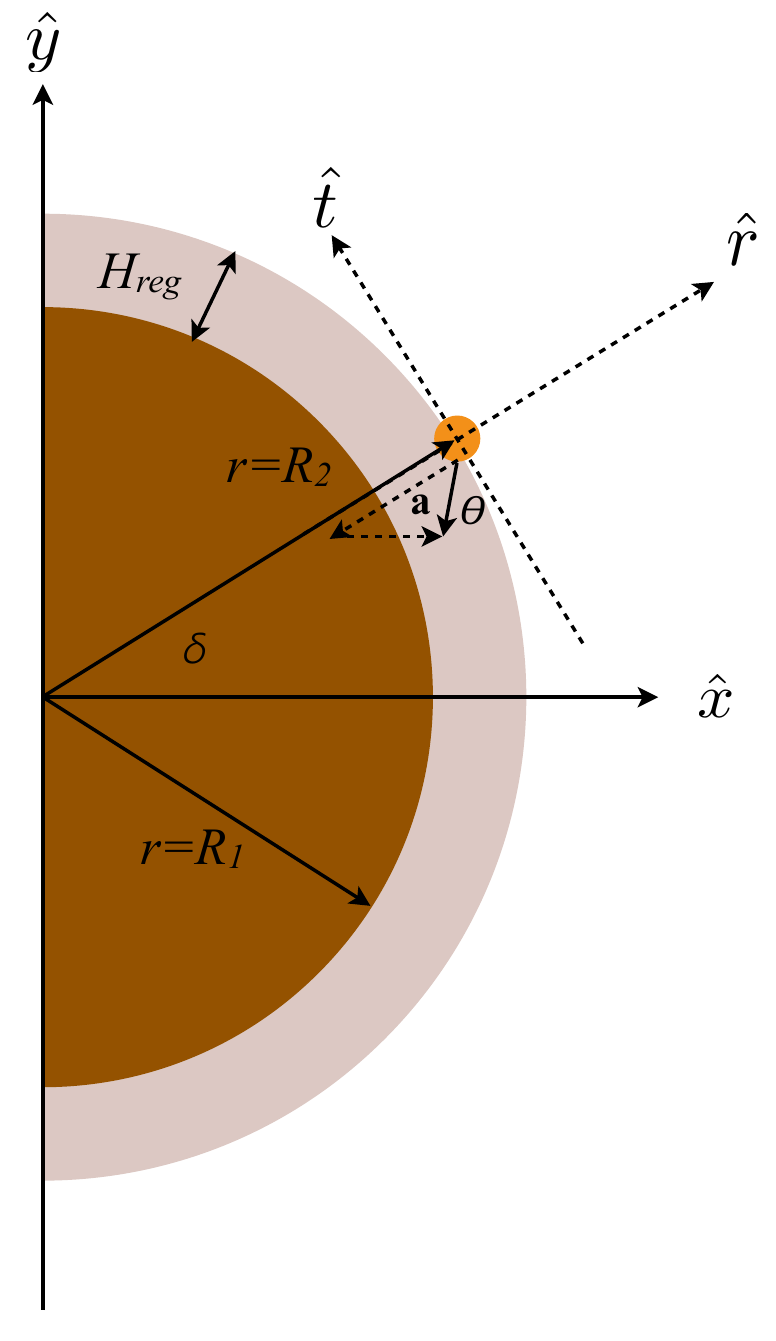}
\caption{Cross-sectional cut of the test body and force diagram of a particle in the shell.}
\label{shell}
\end{center}
\end{figure}

Figure \ref{shell} shows a profile of the test body, indicating that the model asteroid has a solid, monolithic (or relatively strong) core covered by a layer of regolith, similar to \cite{scheeres2015}.  The radius of the test body is $R_2$, the radius of the core is $R_1$ while the thickness of the shell is $H_{reg}$.  In the test body reference frame, a particle in the shell, will feel a gravitational force directed to the center of the test body and a centrifugal force due to rotation.  The latter force changes in magnitude and direction with the latitude $\delta$ of the particle, but is always perpendicular to the axis of rotation whereas the gravitational force is always pointing to the center of the test body and its magnitude depends solely on the distance of the particle to it. It is assumed that the core and the individual regolith particles are made of the same material of density $\rho_1$; however, given that the shell is not monolithic, it will have a filling fraction of $\phi$ and a bulk density $\rho_2 = \phi\rho_1$.  

For this arrangement the gravitational acceleration that a particle in the shell would feel is:
\begin{equation}
{\bf\vec g}=-\frac{4}{3}\pi G\left(\frac{R_1^3(\rho_1-\rho_2)}{r^2}+r\rho_2\right){\bf\hat r}
\label{shell-eq}
\end{equation}
where $G$ is the gravitational constant, $r$ is the distance from the centre of the test body to the centroid of the particle in the shell and all the other variables as are specified before.  The weight of a particle in the shell is then given as $m {\bf\vec g}$, where $m$ is the mass of said particle.
This detailed gravitational model is used in the simulations. For the analytical results the simplifying assumption that the grain is at the surface of the monolithic core is made, and that $H_{reg} \ll R_1$. 

The centrifugal force in the body frame can be projected into two orthogonal components:
\begin{eqnarray}
{\bf \vec f} & = & f_r{\bf\hat r} + f_t \bf\hat t \\
{f}_r&=&mr\omega^2\cos^2(\delta)\\
{f}_t&=& - mr\omega^2\cos(\delta)\sin(\delta)
\end{eqnarray}
where $\omega$ is the rotation rate  in rad s$^{-1}$ of the test body.

Finally, the model of cohesive force acting on a given grain is described. Following the model developed in \cite{sanchez2014}, an overall cohesive strength of the regolith $\sigma_c$ and a characteristic area $\Delta A$ is defined (generally related to the size of the grain), giving a scalar quantity $f_c = \sigma_c \Delta A$. Calculations based on lunar regolith and inferences from rotationally disrupted asteroids have shown that cohesive strength of the regolith is expected to vary between near-zero and 100 Pa \citep{hirabayashi2014c}. 

From a detailed mechanics perspective it can be noted that the tensile strength of the aggregate is related to the particle-particle tensile strength by \citep{sanchez2016}\footnote{There is a typo in this equation in \citep{sanchez2016} as the denominator there has a $\sqrt[3]3$ and it should be $\sqrt{3}$}:
\begin{equation}
\sigma_{yy}^a=\sigma_{yy} \frac{\phi\cos(45^{\circ})}{\sqrt{3}}.
\end{equation}
and that the cohesive strength, or simply cohesion, can be obtained  from the following expression:
\begin{equation}
\sigma_c=\sigma_{yy}^a \tan{\theta} \label{eq:sigmac}
\end{equation}
where  $\theta$ is the angle of friction of the aggregate.  
Referring to the Mohr-Coulomb yield criterion, if cohesion (or cohesive strength) is defined as the shear stress at zero normal stress, then tensile strength is the normal stress at zero shear stress.  
For the current simulations and analytical work, an angle of friction of $35^\circ$ is assumed, in accordance with usual granular media. The filling fraction for the simulations varies locally between $\phi = 0.5 \rightarrow 0.6$, so 0.55 is used for the bulk value. 


\paragraph{Non-dimensional Cohesion} 
A common way to specify the strength of a given cohesive force is to define the Bond number, $B_c$, which is the cohesive force acting on a grain divided by its weight, where the gravitational weight $m g$ is used. 
\begin{eqnarray}
	B_c & = & \sigma_{yy}^a \Delta A / (m g) \label{eq:bond}
\end{eqnarray}
We note that for small asteroids which have a low gravitational acceleration this number can be quite large, even for relatively large grains \cite{scheeres2010}. Take the area of a particle of size $d$ to be $d^2$, and its mass to be $\rho d^3$. The surface gravity is taken as $\sim 4{\cal G}\rho R$. Then the bond number is $\sigma_c / (4{\cal G} \rho^2 d R)$. For a weak regolith $\sigma_c \sim 3$ Pa, a density of $3200$ kg/m$^3$, and ${\cal G} = 6.67\times10^{-11}$ m$^3$/(kg s$^2$). This gives $B_c \sim 1100 / (d R)$. For a millimeter sized grain there is a greater than unity bond number at asteroids smaller than $R \sim 1100$ km. For asteroids less than 1000 m, a millimeter-sized regolith grain will have a bond number greater than 1000, and for bodies less than $\sim 100$ m, bond numbers over 10,000! 

The bond number parameter can also be used to evaluate the relative cohesive strength acting on larger boulders embedded in a regolith of a given cohesive strength. For the same level of cohesion in the regolith and grain density, the size of boulder at which the bond number is unity can be found, yielding 
\begin{eqnarray}
	d & = & 1100 / R
\end{eqnarray}
where $R$ is in meters. Thus, for a 10 km asteroid, decimeter-sized rocks will have unity bond numbers, on a 1 km asteroid meter-sized rocks, and on a 100 m asteroid decameter-sized rocks. 

\section{Failure Theory of Cohesive Surface Regolith}

To understand when surface regolith or granular material on a rapidly spinning asteroid may fail, classical Coulomb theory is used motivated by the sliding block model \cite{nedderman}. Assume the regolith has an angle of friction $\theta$ and is affected by the forces outlined above. The classical theory of Coulomb friction failure is that a slope will fail (i.e., landslide) when the tangential force acting on it exceeds the combined friction and cohesive forces
\begin{eqnarray}
	| f_t |  & \ge & (m g - f_r) \tan\theta + f_c \label{eq:landslide}
\end{eqnarray}
where $m$ is the characteristic mass of a grain. 

This failure law only applies when the grain feels a positive-down effective weight, however, or $mg > f_r$. 
For a system with non-zero cohesion, a landslide failure may not occur up to this point, thus the transition past this condition must also be understood. 
Once the spin rate increases such that $mg - f_r \le 0$ the failure model changes to a tensile failure, what we call fission failure. In this regime failure occurs when the total magnitude of the centrifugal force acting on the grain exceeds the combined force of gravity and cohesion. As cohesion acts against the total force acting on it, the condition becomes
\begin{eqnarray}
	\sqrt{ f_t^2 + (f_r-mg)^2 } & \ge & f_c \label{eq:fission}
\end{eqnarray}

These two failure conditions are evaluated in the following. 

\subsection{Landslide Failure}
First consider failure by landslide, defined by Eqn.\ \ref{eq:landslide}. 
Substituting this model and assuming that $g = \mu /R^2$, where $R$ is the nominal radius of the monolithic body, the condition for failure at a given latitude becomes
\begin{eqnarray}
	\omega^2 R \sin\delta \ \cos\delta & \ge & \left| \frac{\mu}{R^3} - \omega^2 \cos^2\delta\right| R \tan\theta + \sigma_c \Delta A / m
\end{eqnarray}
so long as $\mu / R^3 > \omega^2\cos^2\delta$. 
This condition can be solved for the spin rate at which failure will occur, yielding
\begin{eqnarray}
	\omega^2 & \ge & \frac{\mu}{R^3} \frac{\tan\theta + B_c}{\cos\delta\left(\sin\delta + \cos\delta\tan\theta\right)} \label{eq:fail_flow}
\end{eqnarray}
{\color{black}
where the definitions in Eqns.\ \ref{eq:sigmac} and \ref{eq:bond} are used. 
}
We introduce the non-dimensional spin rate $\Omega = \omega / \sqrt{\mu/R^3}$ and apply 
{\color{black}
the angle sum identity for $\sin(\delta+\theta)$ 
}
to the expression to find the non-dimensional failure criterion for when landsliding will occur
\begin{eqnarray}
	\Omega_L^2 & = & \frac{\left[\sin\theta + \cos\theta B_c\right]}{\cos\delta \sin(\delta+\theta)}
\end{eqnarray}
For a given angle of friction and bond number the minimum spin rate for failure across all latitudes is determined by solving for the value of $\Omega_L$ at which $\partial{\Omega_L^2}/{\partial\delta} = 0$. Doing so yields the minimum spin rate of 
\begin{eqnarray}
	\Omega_L^{*2} & = & \frac{2 \left[\sin\theta + \cos\theta B_c\right]}{1 + \sin(\theta)}
\end{eqnarray}
which occurs at a latitude of $\delta^* = \pi/4 - \theta/2$ and is independent of the Bond number. 

\subsection{Fission Failure}
Now consider failure by fission, defined by Eqn.\ \ref{eq:fission}.  The transition occurs when $f_r \ge m g$, which corresponds to $m R \omega^2 \cos^2\delta \ge m \mu / R^2$. 
In terms of the non-dimensional spin rate the grain will fail under fission once $\Omega^2 \ge \Omega_T^2$ where $\Omega_T = 1/\cos\delta$ is the normalized transition spin rate and is a function of latitude.

Introducing the previously defined non-dimensional parameters, Eqn.\ \ref{eq:fission} can be reduced to 
\begin{eqnarray}
	\Omega^4 - 2\Omega^2 + \frac{1-B_c^2}{\cos^2\delta} & \ge & 0
\end{eqnarray}
The roots of this equation can be resolved, and the fission failure condition that exceeds the transition spin rate is 
\begin{eqnarray}
	\Omega_F^2 & = & 1 + \sqrt{\frac{B_c^2 - \sin^2\delta}{\cos^2\delta}} , 
\end{eqnarray}
whereas the other root occurs at less than the transition rate, and thus is not relevant. 
We can also evaluate this limiting spin rate for its minimum values. Computing the partial and evaluating shows that at the equator ($\delta=0$) the spin rate will locally be maximum for $B_c < 1$ and a minimum for $B_c > 1$, with the failure spin rate being constant across all latitudes and equal to $\sqrt{2}$ at $B_c=1$. 

\subsection{Failure Phase Diagram}

An important consideration is when the surface will fail by landslide and when it will fail by fission (fail in tension). To find this limit, equate the transition spin rate to the landslide spin rate, $\Omega_T^2 = \Omega_L^2$ and evaluate the level of cohesion necessary to sustain this rate of spin (we can conversely also solve $\Omega_T^2 = \Omega_F^2$ to find the same limit). Then the limiting Bond number $B_c^*$ is a function of latitude but independent of surface friction and is
\begin{eqnarray}
	B_c^* & = & \tan\delta
\end{eqnarray}
Thus, at a given latitude of the body, if $B_c < \tan\delta$, that portion of the surface will fail by landslide, while if $B_c > \tan\delta$ it will fail in tension. 
This allows one to map out the landscape of failure conditions as a function of cohesive bond number (see Fig.\ \ref{fig:bondlim}). Regions of the surface with Bond number below the curve will fail by landslide, while those above will fail in tension. Where the two failure regimes touch we find more complex failure dynamics. 

\subsection{Orbit Elements of Failed Regolith} 

Once a regolith grain fails, especially if it loses contact with the surface, then it will enter an orbital regime. Thus, it is relevant to determine what that initial orbit will be. It can be noted that grains that fail by landslide do not immediately enter orbit, and their migration patterns were discussed in \cite{scheeres2015}, where it was noted that they may enter orbit once they migrate down to the equatorial region. Thus, the current discussion is mainly focused on the fission of regolith grains from the surface. 

Once the cohesive force between the surface (or cohesive matrix) and grain is broken, the centrifugal forces transform to an instantaneous acceleration relative to inertial space and the gravitational force remains active. The initial velocity of the particles are parallel to the equatorial plane and perpendicular to the radius vector, with a speed $V = \omega R \cos\delta$. 
{\color{black}
This occurs as the particle is stationary on the surface and has a position vector ${\bf R} = R\cos\delta\hat{\bf r} + R\sin\delta\hat{\bf z}$. Then the velocity vector is ${\omega}\hat{\bf z}\times {\bf R}$. 
}
By definition the grain is initially at periapsis, so
\begin{eqnarray}
	R & = & a (1-e)
\end{eqnarray}
where $a$ is the grain orbit semi-major axis and $e$ is its eccentricity. The angular momentum of the grain is $H = RV$ and thus
\begin{eqnarray}
	\sqrt{\mu a(1-e^2)} & = & \omega R^2 \cos\delta
\end{eqnarray}
From this the eccentricity and semi-major axis of the grain orbit can be found
\begin{eqnarray}
	e & = & \left(\frac{\Omega}{\Omega_T}\right)^2 - 1 \\
	a & = & \frac{R}{2-\left(\frac{\Omega}{\Omega_T}\right)^2}
\end{eqnarray}
where $\Omega_T = 1 / \cos\delta$. 
First, from the condition $e \ge 0$ it is seen that direct entry into orbit only occurs when $\Omega \ge \Omega_T$, which places failure in the fission regime, as noted above. 

The energy of the orbit can also be computed, and is found to be
\begin{eqnarray}
	E & = & - \frac{\mu}{R} \left[ 1 - \frac{1}{2} \left(\frac{\Omega}{\Omega_T}\right)^2\right] 
\end{eqnarray}
Thus for $1 \le \left(\Omega / \Omega_T\right)^2 \le 2$ the orbital energy is negative, meaning that the orbit is bound and that the grain will ideally return to the surface after one orbit period. The maximum distance that the grain will travel from the surface will be the radius of apoapsis, $R_A = a(1+e)$, and equal to 
\begin{eqnarray}
	R_A & = & \frac{R \ \left(\frac{\Omega}{\Omega_T}\right)^2}{2-\left(\frac{\Omega}{\Omega_T}\right)^2}
\end{eqnarray}
This condition can be related to the magnitude of the Bond number by resolving the inequality $\Omega_T^2 \le \Omega_F^2 \le 2 \Omega_T^2$, yielding  
\begin{eqnarray}
	\tan\delta & \le B_c \le & \sqrt{ 1 + 4 \tan^2\delta} 
\end{eqnarray}
Thus if $B_c$ falls in this limit, the initial orbit is bound and the regolith grain will return to its periapsis at the surface, unless additional orbital perturbations either raise its periapsis, placing it in orbit about the body, or reduce its periapsis making it reimpact. 

\begin{figure}[ht!]
\centering
\includegraphics[scale=0.25]{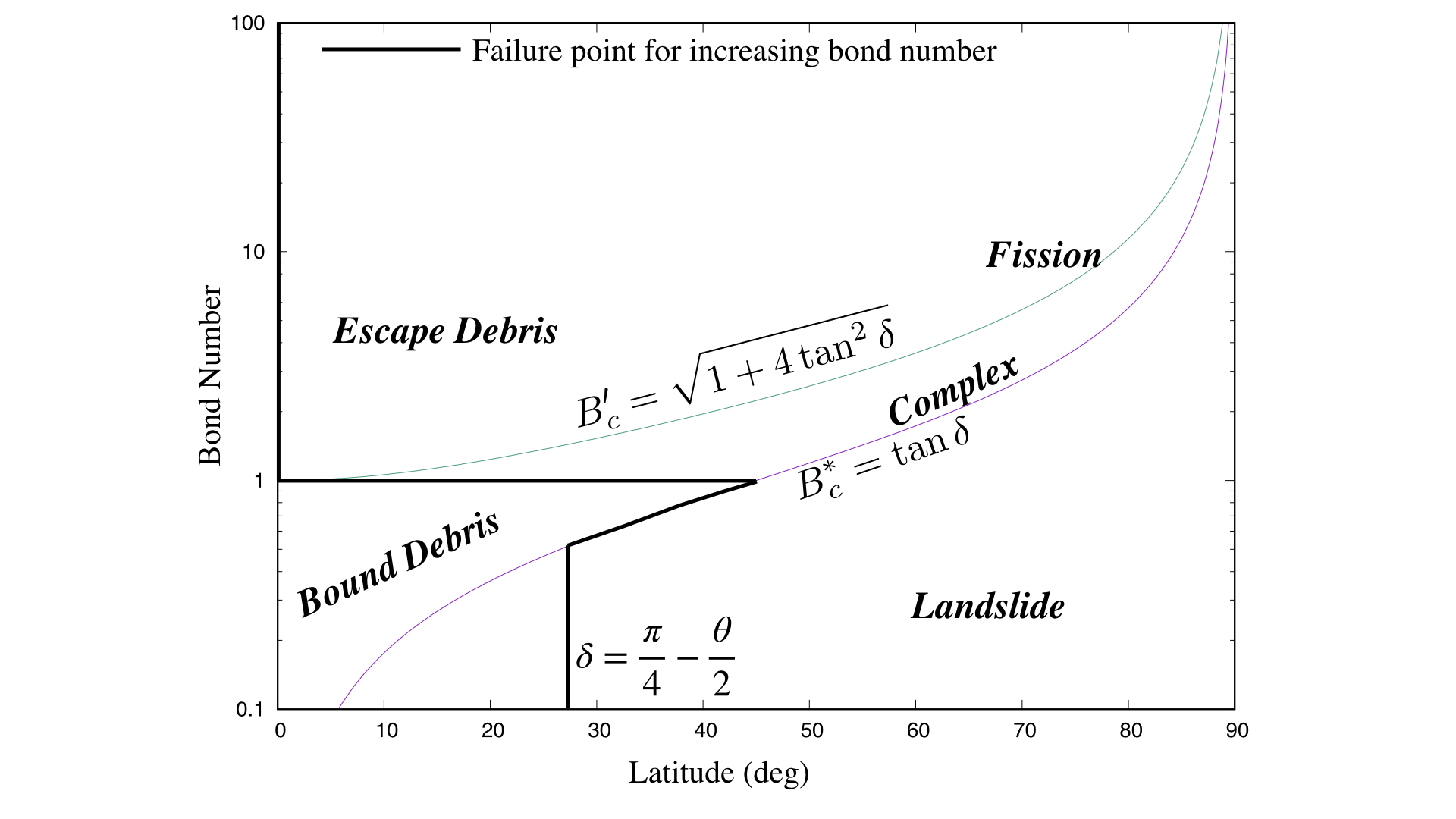}
\caption{Failure type phase diagram as a function of surface latitude and cohesive bond number. The heavy black line traces the latitude at which failure occurs first as a function of Bond number, here for a friction angle of $35^\circ$. The purple line is the dividing line between landslide and fission, and is defined as ``complex'' as it involves both mechanisms in general. The green line lies in the fission region and indicates the dividing line when fissioned material will escape (above the line) and when it will initially be bound (below the line). }
\label{fig:bondlim}
\end{figure}

For Bond numbers greater than or equal to $\sqrt{1+4\tan^2\delta}$ the orbital energy will be positive and the grain will be directly placed on a parabolic or hyperbolic orbit, taking it away from the parent body and reducing the mass of the system. These Bond number limits are shown in Fig.\ \ref{fig:bondlim}. 

In addition to the semi-major axis and eccentricity, the other orbital elements can also be defined for the fissioned particle. The inclination of the orbit will equal the absolute value of the fission latitude, $i = |\delta|$. The longitude of the ascending node will be $90^\circ$ before the longitude of the particle if the latitude is positive, and $90^\circ$ after the longitude if the latitude is negative. 
The argument of periapsis will equal $\pm90^\circ$ depending on whether the latitude is positive or negative. Due to this geometry, for the bound orbits the apoapsis will occur at a latitude of $\mp\delta$, and for the unbound orbits the escape trajectories will be in the opposite North-South hemisphere from which the grain detached. 

\subsection{Failure Spin Rates and Latitudes}

\begin{figure}[ht!]
\centering
\includegraphics[scale=0.25]{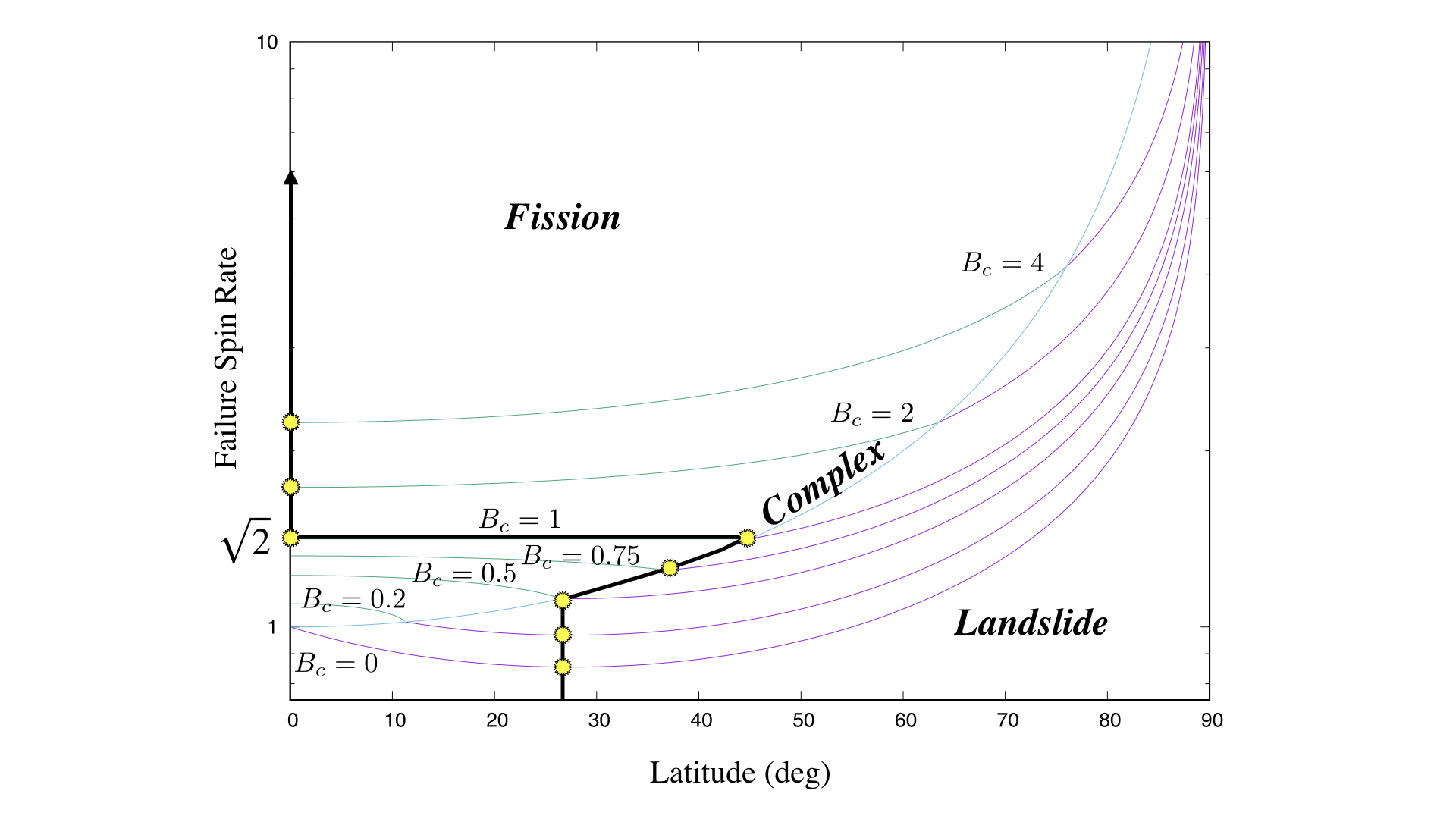}
\caption{Failure spin rate as a function of latitude for different cohesive bond numbers. The yellow dots signify the latitude where the first failure occurs for a given bond number and the black line traces the point of first failure for increasing bond numbers. The blue line is the dividing line between landslide and fission, and is classified as ``complex'' failure as it involves both mechanisms in general. The purple lines are landslide failure conditions at constant bond number, while the green are fission failure conditions at constant bond number.  }
\label{fig:spinlim}
\end{figure}

The spin rate at which failure occurs will vary across the surface. Figure \ref{fig:spinlim} shows the failure spin rates as a function of latitude for different cohesive Bond numbers. First note that the migration of the first failure point as the Bond number increases (also indicated on Fig.\ \ref{fig:bondlim}). In addition to this there are several interesting aspects of this graph as the Bond number changes, described below. 

\paragraph{$\bf B_c = 0$}
First consider the cohesionless case. Here note that failure always occurs via landslides. Further, this failure always occurs first at $\delta^* = \pi/4 - \theta/2$ at a value $\Omega = \sqrt{\frac{2 \sin\theta}{1+\sin\theta}} < 1$.  As spin rate increases beyond this point, the failure region spreads to lower and higher latitudes. Eventually, once $\Omega = 1$ the failure zone reaches the equator and at the other end a latitude of $\pi/2 - \theta$. For increasing spin rate the failure latitude continues to grow to higher latitudes, although only reaches the pole as $\Omega \rightarrow \infty$. Note that in this paper only the failure of a relatively thin covering of regolith is considered, unlike the analysis of cohesionless regolith in \cite{scheeres2015} which also accounted for significant depth. 

\paragraph{$\bf 0 < B_c < \tan(\pi/4 - \theta/2)$}
In this range failure first occurs by landslide at a latitude of $\delta^* = \pi/4-\theta/2$, equal to the cohesionless case. The failure spin rate is now $\Omega = \sqrt{\frac{2 \left( \sin\theta + \cos\theta B_c\right)}{1+\sin\theta}}$ and is still less than unity. If the spin rate is increased from this value, failure by landslide occurs at higher and lower latitudes, up to the spin rate $\sqrt{1+B_c^2}$, at which point the lower latitude at $\delta = \arctan(B)$ will transition into failure by fission. This continues up to a spin rate of $\sqrt{1+B_c}$ at which the equator also fails by fission. At higher latitudes, the landslide failure point migrates to larger values, again asymptotically reaching the poles only when $\Omega \rightarrow \infty$.  

 \paragraph{$\bf \tan(\pi/4-\theta/2) < B_c < 1$}
 In this range the surface will fail first at a latitude of $\delta^* = \arctan(B_c)$, which occurs at the transition point between landsliding and fission. As $B_c$ approaches unity the failure latitude increases towards $\pi/4$. The initial failure spin rate is $\sqrt{1+B_c^2}$, and as the spin rate is increased beyond this value the lower latitudes will fail by fission, while the upper latitudes by landsliding. The limiting spin rates at the equator and the poles are similar to those given above. Since the fission and landsliding failure regimes are directly next to each other the failure dynamics are more complex, and can involve features of both types of failure. This is discussed more in the simulation section. 

\paragraph{$\bf B_c = 1$}
At a unity bond number there is a fundamental transition in failure modes, where the entire body surface up to a latitude of $45^\circ$ will fail first in tension at a spin rate of $\Omega = \sqrt{2}$. At higher spin rates failure will occur in landslide at higher latitudes. Note that at this spin rate, any regolith that is separated from the surface at the equator will immediately have escape speed, and thus this cohesive limit also demarcates when an asteroid would enter its disaggregation phase \cite{scheeres2018}. However, the failed regolith at higher latitudes will still be initially bound to the asteroid and most likely return and impact later. 

\paragraph{$\bf B_c > 1$}
In this realm initial failure always occurs at the equator at a spin rate of $\sqrt{1+B_c}$. For increasing spin rate the failure by fission will move to higher latitudes. The fissioned particles will immediately escape up to a latitude of $\delta = \arctan( \sqrt{(B_c-1)/2} )$, beyond which the fissioned particles will be on bound orbits. 
Once the spin rate goes beyond $\sqrt{1+B_c^2}$, failure will occur by landsliding again, with this transition occurring at a latitude of $\delta = \arctan(B_c)$. 

\section{Regolith Failure Simulation Model}

Given the theoretical analysis of failure as a function of latitude, cohesion and spin rate, detailed granular mechanics simulations are carried out to explore the theory and to validate some of the basic conclusions given above. The following describes the methodology used and presents a number of detailed simulations to better understand surface regolith failure. An important aspect of the model is that, through using non-dimensional models, it can use one set of simulations to study a range of different physical systems. 

\subsection{Numerical Modeling Details}
First the details of the numerical model are described. 

\paragraph{Geometry of the Simulations}
To avoid simulating an entire spherical body covered with grains, a lune of angular width $\Delta\lambda$ is constructed within which all of the simulations are carried out. The physical width of the lune will then be $\Delta W = R \cos\delta \Delta\lambda$, shrinking to an ideal point at the poles. 

\begin{figure*}[h]
\begin{center}
\includegraphics[max size={\textwidth}{\textheight}]{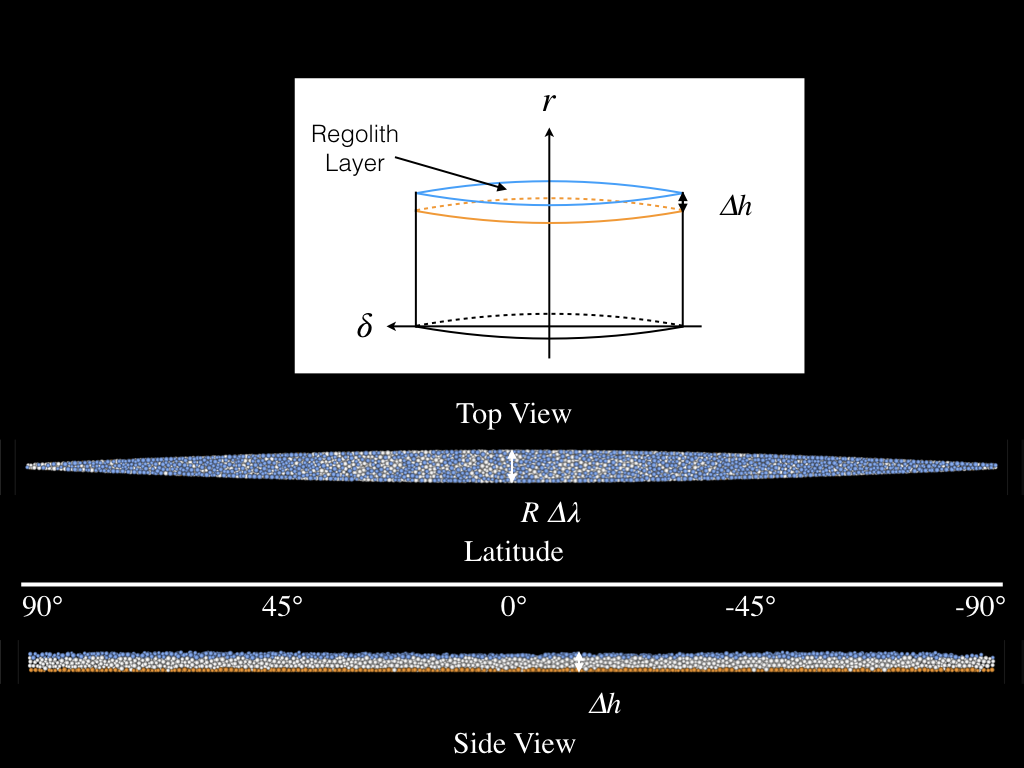}
\caption{Top: Test body lune, mapped into a cartesian coordinate system. Bottom: Initial configuration for all simulations.}
\label{sim-ini}
\end{center}
\end{figure*}

This lune can be mapped into a Cartesian coordinate system with axes $\bf\hat r$ and $\bf\hat t$, with the coordinate along $\bf\hat t$ being the latitude $\delta$ as is depicted in fig.~\ref{sim-ini}.
From this mapped lune, only the uppermost regolith layer is simulated, as this is the focus of this study.

\paragraph{Granular Mechanics Model}

The simulation program that is used for this research applies a Soft-Sphere Discrete Element Method (SSDEM) \citep{cundall1971, cundall1992}, implemented as a computational code (in house developed) to simulate a granular aggregate \citep{bis, sanchez-lpsc2009, sanchez2011,sanchez2012}. The particles, modeled as spheres that follow a predetermined size distribution, interact through a soft-repulsive potential when in contact.  This method considers that two particles are in contact when they overlap.  When this happens, normal and tangential contact forces are calculated \citep{herr1}. The former is modeled by a hertzian spring-dashpot system and is always repulsive, keeping the particles apart; the latter is also modeled with a linear spring that satisfies the local Coulomb yield criterion.  The normal elastic force is modeled as
\begin{equation}
{\vec{\bf f}}_e= k_n\xi^{3/2}{\bf\hat n},
\label{hook}
\end{equation}
the damping force as:
\begin{equation}
{\vec{\bf f}}_d=-\gamma_n\dot\xi{\bf\hat n},
\end{equation}
and the cohesive force between the particles is calculated as
\begin{equation}
{\vec{\bf f}}_c=-2 \pi \frac{r_1^2 r_2^2}{r_1^2 + r_2^2} \sigma_{yy} \hat{\bf{n}}
\end{equation}
where $r_1$ and $r_2$ are the radii of the two particles in contact, $\sigma_{yy}$ is the tensile strength of this contact, which is given by a cohesive matrix formed by the (non simulated) interstitial regolith \citep{sanchez2014}, and $\hat{\bf{r}}_{12}$ is the branch vector between the centres of these two particles.  Then, the total normal force is calculated as ${\vec{\bf f}}_n={\vec{\bf f}}_e+{\vec{\bf f}}_c+{\vec{\bf f}}_d$.  In these equations, $k_n$ is the elastic constant, $\xi$ is the overlap of the particles, $\gamma_n$ is the damping constant (related to the dashpot), $\dot\xi$ is the rate of deformation and ${\bf\hat n}$ is the vector joining the centres of the colliding particles. This dashpot models the energy dissipation that occurs during a real collision.

The tangential component of the contact force models surface friction statically and dynamically. This is calculated by placing a linear spring attached to both particles at the contact point at the beginning of the collision \citep{herr1,silbert2001} and by producing a restoring frictional force ${\vec{\bf f}}_{t}$. The magnitude of the elongation of this tangential spring is truncated in order to satisfy the local Coulomb yield criterion $|{\vec{\bf f}}_t|\leq\mu |{\vec{\bf f}}_n|$. 

Rolling friction \citep{ai-chen2011,hirabayashi2015,sanchez2016} has also been implemented in order to mimic the behaviour of aggregates formed by non-spherical grains.  Particles are subjected to a torque that opposes the relative rotation of any two particles in contact.  This torque, similar to surface-surface friction, is implemented as linearly dependent on the relative angular displacement of any two particles in contact.  {\color{black}At every timestep, the relative angular displacement is calculated as: $\Delta M^k_r=-k_r \Delta\theta_r$, where $k_r$is the rolling stiffness and $\theta_r$ is their incremental relative rotation.

 The rolling resistance torque }has a limiting value of:
\begin{equation}
M_r^m=\mu_r R_r |{\vec{\bf f}}_n|
\end{equation}
where $\mu_r$ is the coefficient of rolling resistance, $R_r=r_1r_2/(r_1+r_2)$ is the rolling radius, and $r_1$ and $r_2$ are the radii of the two particles in contact .  This allows the simulations to reach angles of friction of up to $\approx 35^{\rm o}$ as evaluated by the Druker-Prager yield criterion \citep{sanchez2012}.  This value for the angle of friction is typical of  {\color{black}cohesionless granular aggregates}, though friction angles of $\sim$40$\rm^o$ are not rare. This implementation of rolling friction is similar to that of surface friction, but is instead related to the relative angular displacement.  

\paragraph{Simulation Details}

In addition to these contact forces, gravitational and centrifugal forces, as well as the specific geometry of the lune have also been implemented in the code so that the simulation is faithful to the model.  Fig~\ref{sim-ini} shows the initial setup for all simulations.  The uppermost layer of particles have been colored in blue to facilitate visualization.  The white and blue particles are completely free to move under the influence of the forces in the simulation.  The orange particles at the bottom are stuck to the bottom of the simulation box to simulate the rough, solid surface of the core of the test-body.  The walls containing the particles, which also provide the granular medium its shape (not depicted) are completely frictionless.  {\color{black}  Experiments carried out in an inclined channel with a flow-rate-controlled system have shown that for high flow rates, the flow occurs atop a static granular heap whose angle is considerably higher than those usually exhibited by granular heaps \cite{taberlet2003}.  The unusual stability of these heaps can be accounted for by the flowing layer and its friction on the sidewalls \cite{taberlet2008}.  The walls were made frictionless in order to avoid this artificial strengthening of the aggregates \cite{richard}.}  All simulations were carried out with 5000 spherical particles with a density of 3200 kg m$^{-3}$; the particle diameters follow a uniform distribution and were randomly selected in a range between 2-3 cm.  This arrangement provided a regolith layer of $\approx$ 12 cm in a rectangular simulation box that was 5.3 m in length.  This means that the test body being simulated has a diameter of 3.4 m.  If the height of the regolith layer we increased, the size of the core would be reduced so that the size of the test body is kept constant; this would also imply the recalculation of the  gravitational field. Note that the theory allows one to rescale the results of the simulations to other sized bodies, once the effective Bond number is defined. 

Initially, the particles are placed inside the rectangular simulation box in a hexagonal closed packed lattice and given random initial velocities.  {\color{black}The code is such that the minimum distance between any two neighbours is 2.1 times the size of the largest particle in the system.  This will provide the particles with enough space to move and reset the initial, ordered configuration.}  Then the rounded walls approach from the outside until they touch on both, left and right ends of the box and the geometry of the stretched lune is reached; then the settling process begins.  Given that the equations have not been normalized, the system has to settle to the ambient gravity that an asteroid of the size and characteristics of the test body would produce which is in the order of 10$^{-6}$ m s$^{-2}$.  To do this, the procedure detailed in \cite{sanchez2011} is followed; this is, the particles (frictionless and cohesionless) are first settled under terrestrial conditions and then gravity is reduced by one order of magnitude until the system resettles.  As the walls of the lune are narrower near both ends of the box, the initially settled particles do not have a flat surface {\color{black}(a valley is formed at the centre)}.  To correct this, the top of the box was lowered until it is as high as the highest particle and then the particles are again given random initial velocities.  Before the next reduction in gravitational field, the maximum  {\color{black}height} of the particles is recalculated so that the top of the box can be lowered again. {\color{black}Given the higher density of particles near the ends, they will naturally tend to migrate towards the centre and the flat top with which they collide will provide a flat surface that will be mimicked.}  This procedure is repeated until the order of magnitude of the desired gravitational field is reached.  Once this happens, the system is settled one last time to its exact gravitational field.  The excess energy that is liberated from the springs as gravity is  {\color{black}reduced is removed} by adding a Stokes' like drag to the particles which speeds up the settling process \citep{sanchez}.  

After the particles  settled in the correct geometry, the top of the box is repositioned to its original height and the spin rate is elevated by 0.1$\omega_c$, where $\omega_c$ is the critical spin rate calculated for failure, every 10 seconds for the cohesive aggregates.  For the cohesionless case, this is changed to 100 seconds when $\omega=0.9\omega_c$.  Spin up is stopped once $\omega_c$ {\color{black} is reached} or any particle of the uppermost layer moved upwards by one particle diameter, but the simulation is continued so that failure can be observed.  {\color{black}Part of the objective of this work is to understand how the regolith on the surface of this idealised body would fail at the critical spin rate and so these simulations were meant to find it.  The simulations take these large jumps because we are exploring where failure happens.  There is no point in exploring smaller intervals in a range of spin rates at which no increased particle motion is observed.  Once we found a smaller range of values to test, the intervals became smaller and instead of having one simulation in which the spin rate was continuously increased, we had several in which the spin rate was increased to different points in the narrower rage and left there for up to 30000 seconds.  We did this in order to verify that failure was not observed because the condition was not met yet and not because we didn't wait for long enough.  The initial velocity of the particles when the aggregates fail is in the microns per second and so the reason for this long waiting period.}

\section{Simulation Results}

The theory as derived above has a number of specific tests that our simulation capability can model. This can be viewed as a way to verify the realism of the simulation, and conversely can also validate the theory and its predictions. Now consider the predictions in the different regimes. Due to the stochastic nature of the simulations they were averaged over several runs in some cases to gain a better understanding.  {\color{black}Averaging is necessary when the spin rate is already found and the failure mechanism of the aggregates is the focus.  Due to the length of the simulations, about 2 weeks for each one, only 6 simulations at a given spin rate were carried out.}  Also, it is found that the theory is not as precise in predicting what the bond number of regolith is, however the regolith can still be placed within certain ranges of bond number. 

\subsection{Failure for $B_c = 0$}

\begin{figure}[h!]
\begin{center}
\subfloat[]{\includegraphics[max size={0.5\textwidth}{\textheight}]{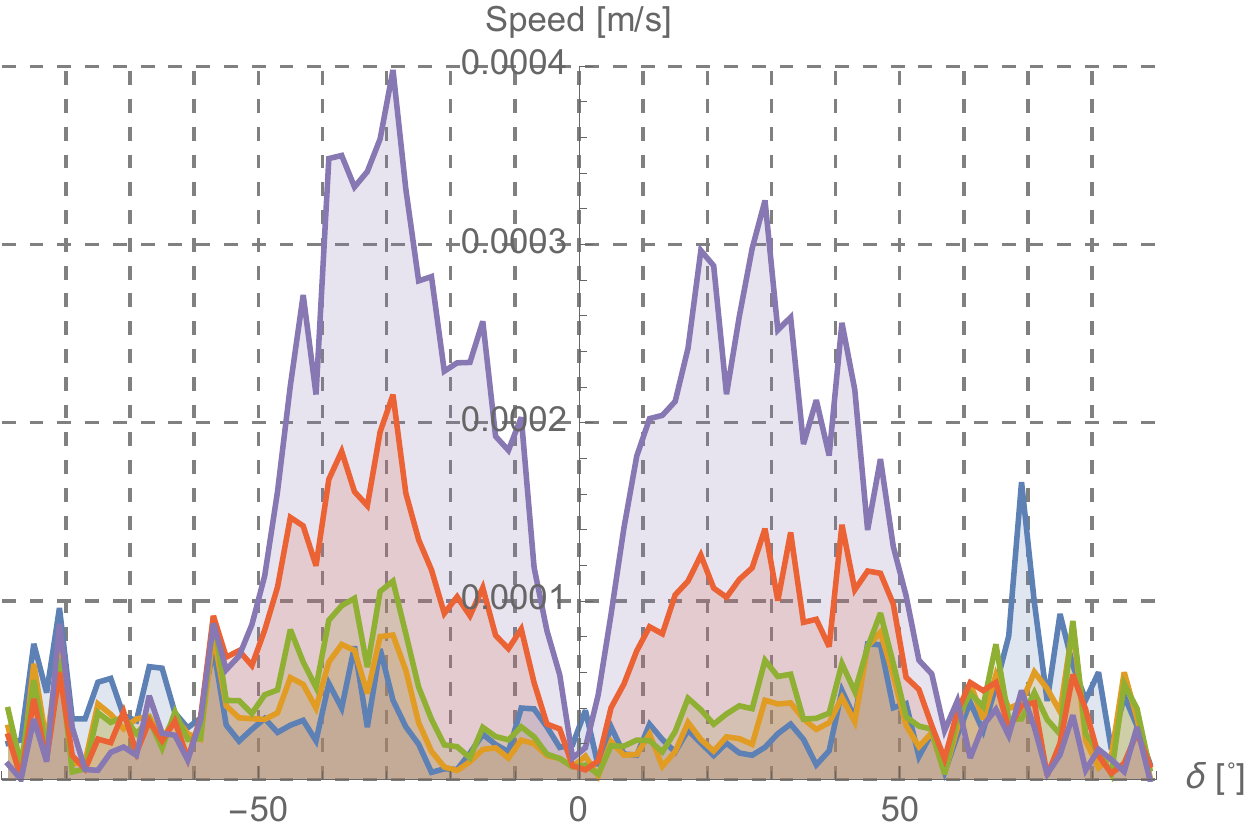}}\\
\subfloat[t=2000 s]{\includegraphics[max size={0.5\textwidth}{\textheight}]{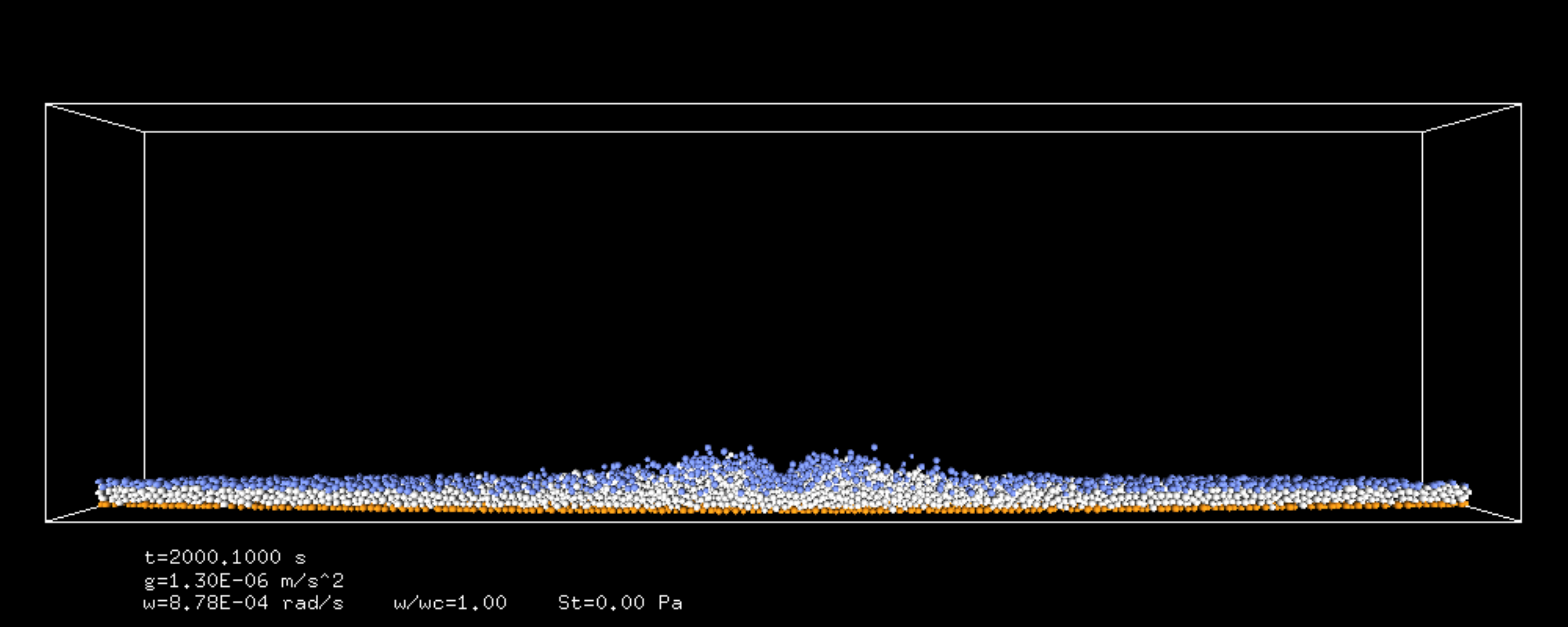}}\\
\subfloat[t=3400 s]{\includegraphics[max size={0.5\textwidth}{\textheight}]{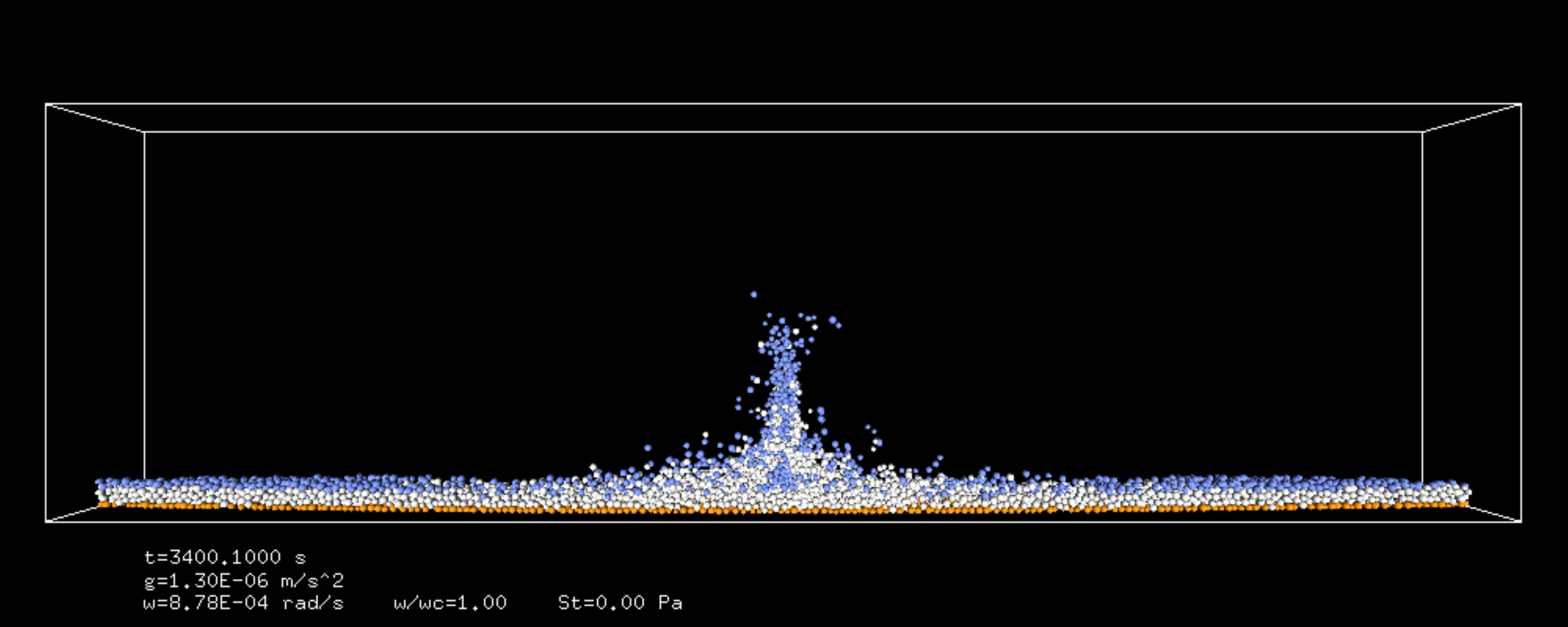}}\\
\subfloat[t=9900 s]{\includegraphics[max size={0.5\textwidth}{\textheight}]{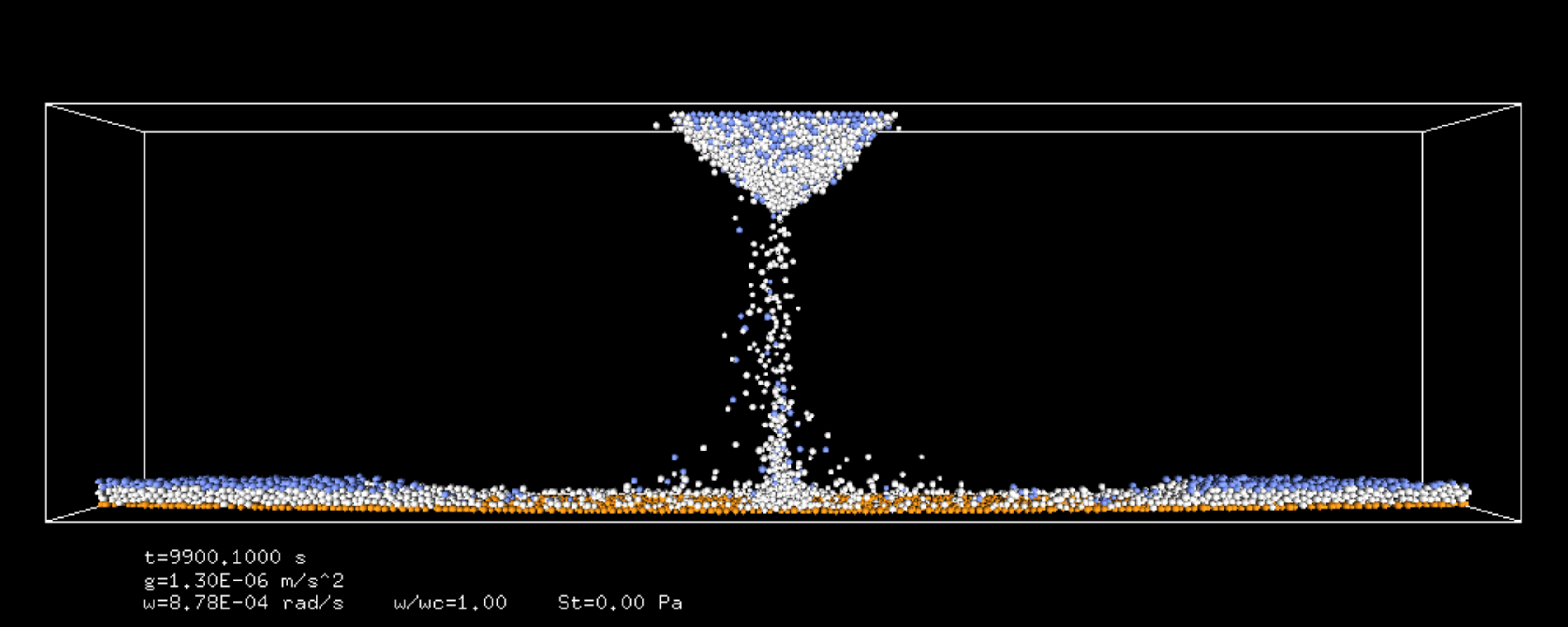}}
\caption{(a)Average speeds of the particles in a cohesionless lune a t = 80, 100, 105, 120 and 150 seconds.  {\color{black} Spin rates are 0.8$\omega_c$ for t=80s and $\omega_c$ from then on.}  Data points are taken in boxes of 3$^{\circ}$ arcs.  (b), (c) and (d) show the motion of the particles at t = 2000, 3400 and 9900 seconds.}
\label{Combined-0Pa}
\end{center}
\end{figure}

One of the predictions of the Coulomb theory was that a cohesionless regolith would first fail at 27.5$^\circ$ if the angle of friction of the grains is 35$^{\circ}$.  Fig.~\ref{Combined-0Pa}(a) shows the average speed of the particles (not taking into account the z component) for t= 80, 100, 105, 120 and 150 seconds.  Each data point was obtained dividing the the simulation box in 60 smaller boxes and averaging the speeds of the particles in each box; therefore, each box represents an arc of 3$^{\circ}$.  This plot evidences two peaks, one above and one below the equator between 27$^{\circ}$-30$^{\circ}$, which is in agreement with the theory.  Notice that the motion of the particles in the simulation {\color{black}is in the order of 10$^{-4}$ to 10$^{-5}$ m/s}.

Figures \ref{Combined-0Pa}(c-d) show the time evolution of the system after $\omega_c$ was reached.  As it can be observed, though the regolith fails at the predicted latitudes, this does not mean the higher and lower latitudes stay intact.  Without the support of neighboring particles and with the added rotation, regolith beyond these latitudes also fails.  In spite of this, particles at the poles ($\approx$ 50$^\circ$ north and south) are left intact at $\omega_c$.  As explained in the previous section, the spin-up process was stopped when $\omega_c$ was reached; however, it is logical that spin rates above it would have resulted in particles closer to the poles being ejected.  

\subsection{Failure for $B_c \ll 1$}

As the bond number is increased to a small value, the failure latitude remained in the vicinity of the predicted $27.5^\circ$ value. This was observed by simulating runs with bond numbers chosen in the range $1\times10^{-4} \rightarrow 1\times10^{-3}$. The failure mode appears similar to the cohesionless case. 

\subsection{Failure for $\tan\delta^* \le B_c \le 1 $}

As the bond number approaches the transition line the onset and subsequent motion of the particles becomes more complex. This is seen as landsliding particles will enter a fission phase once they move down to a lower latitude, leading to a significantly different pattern of failure. This can be seen in a different behavior as shown in Fig.\ \ref{fig:examples}, where the grains start to flow yet are immediately lofted above the surface. 

\subsection{Failure for $B_c \ge 1$}

\begin{figure}[ht!]
\centering
\includegraphics[scale=0.25]{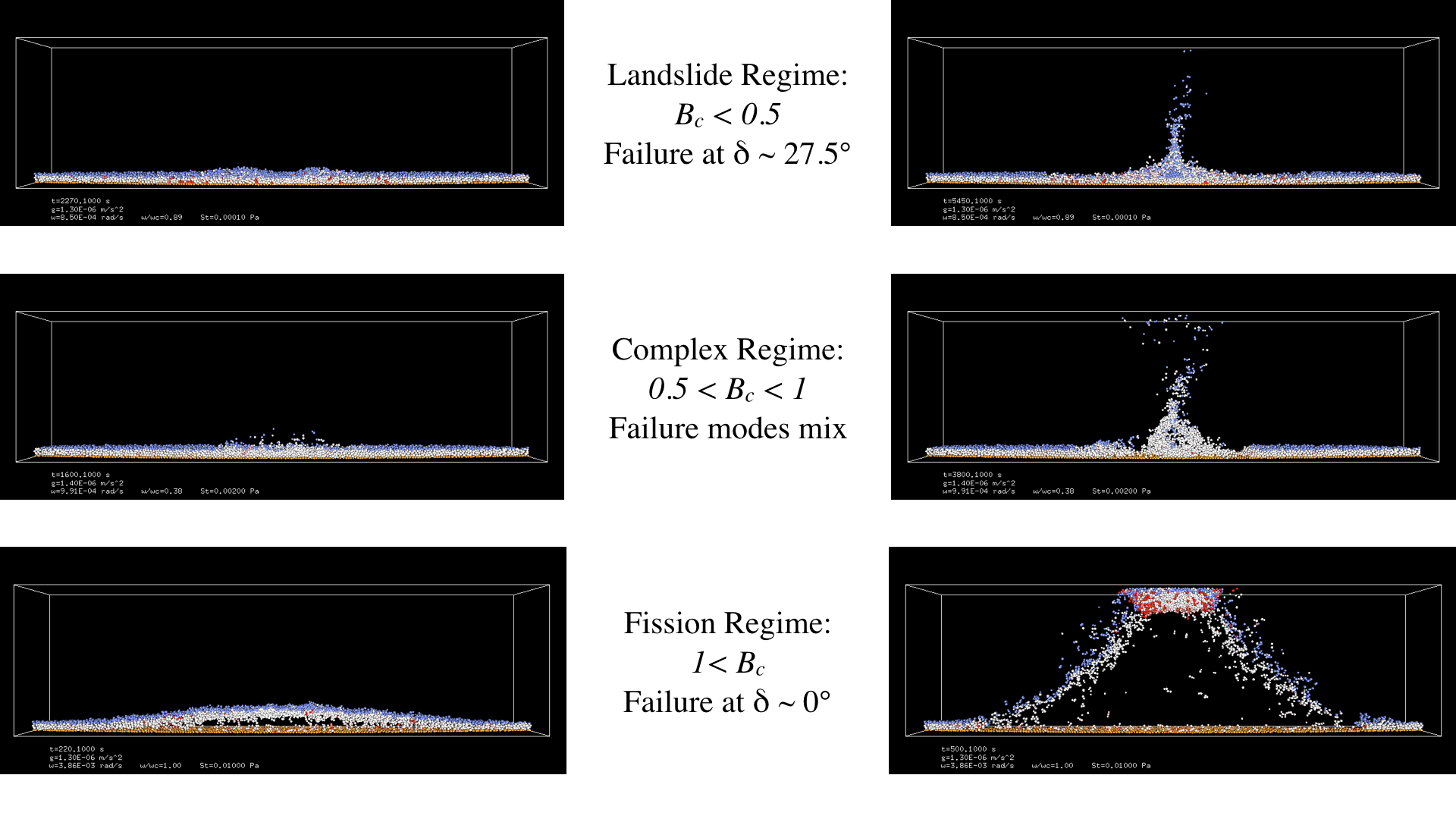}
\caption{Simulations showing different failure patterns as a function of increasing bond number. 
{\color{black}
Figures on the left show the system at start of failure, and on the right show the system after majority of failure has occurred at the given spin rate. 
}
}
\label{fig:examples}
\end{figure}

Once the bond number becomes large enough the initial failure point occurs at the equator. As can be clearly seen in Fig.\ \ref{fig:examples}, due to the modeled cohesion being greater, neighboring grains stay connected and cause the surface to be stripped off of the sub-surface of the asteroid. Note that the ``rind'' of the asteroid is still retained at higher latitudes, and remains resistant to failure as the spin rate is further increased.  

%
%
%

It must also be pointed out that the less cohesive  aggregates were most likely to fail due to small perturbances in the system.  Given the extremely low gravitational conditions, the escape velocity is in the range of $\approx$ 2 mm s$^{-1}$.  This speed could  have been obtained by a particle of average size only by an excessive overlap of only about 4 $\mu$m.  Furthermore, even the strongest tensile strength tested (100 Pa) provides a net force of only 5$\times$10$^{-4}$ N between two average particles.  What this would mean for a SFR is that the slightest collision with a small pebble, if sufficiently energetic, could result in the complete loss of any superficial regolith.

\section{Discussion}

The main question that to be addressed is whether or not a small, super-fast rotator could retain a regolith cover at high spin rates.  However, a more nuanced question that is also 
{\color{black}
addressed
}
here is what are the rotational conditions under which a cohesive regolith layer can be retained, partially lost or completely lost.  If this regolith layer is only partially lost, it would be interesting to know which regions of the test body, and by extension a small asteroid, could potentially be covered.  In order to find answers, we have used the above described simulation method and set up, and have ran simulations with granular systems with particle-particle tensile strength values of that cover the range of bond number failure types.  All of these values are well within the accepted values of tensile strength for observed small NEOs \citep{holsapple2004, holsapple2010, scheeres2010, sanchez2014, rozitis2014, hirabayashi2014c, hirabayashi2015c}.  In general we find that the polar regions of a rapidly rotating body will preferentially retain loose regolith, even in the cohesionless case
{\color{black}
as previously pointed out in \cite{scheeres2015,yu2018}
}
. 

As the derived conditions are couched in non-dimensional terms, the theory can also be applied to bodies of arbitrary shape -- so long as the appropriate non-dimensional parameters are used. For granular mechanics these non-dimensional parameters are the angle of friction and the bond number. Results relating to the angle of friction alone have been shown previously for asteroids, however the inclusion of the bond number into the failure conditions is new in this paper, and highlights important transitions that may occur. One exciting area of application is to the failure conditions for larger boulders on rubble pile bodies, such as analyzed in \cite{tardivel2018}. Future analysis will test the models developed here for their applicability to the loss of larger boulders on rapidly rotating asteroids. 

This study has used an idealized spherical test body in its analysis and simulation. It is understood that such shapes are not found naturally and that even bodies with symmetric shapes have significant deviations from the sphere.  
{\color{black}
Nonetheless, the failure conditions found in the current paper can be generalized to arbitrary bodies by defining a local ``equivalent'' latitude on a body. This would consist of identifying the local slope on a more arbitrary surface and identifying it with the equivalent latitude with that slope. Such applications can be explored in the future. 
}
Further, the general trend of the polar regions being stable is expected to persist even on non-spherical asteroids (as indicated in \cite{yu2018}). This is so as in the polar region of a fast-spinning asteroids the lateral accelerations will still be near-zero and vanish at some point. Thus, so long as the local surface slope at the region is stable, then material is expected to remain there. 

A strong assumption made in this analysis is that of uniform rotation about a fixed axis. Should an asteroid undergo a period of tumbling then this assumption is violated. While if this occurs when the body has a low overall angular momentum, then the centrifugal accelerations will be small and potentially negligible (such as on Toutatis). However, there are small bodies that are seen to be rapidly rotating and tumbling. In this situation the polar regions may be cleared of loose regolith as the ideal zero lateral acceleration conditions will be violated. 

Of interest to test this theory will be spectroscopic and thermal IR observations of fast-spinning asteroids. Comparisons of observations of their polar regions and equatorial regions could show heterogeneity, which could be interpreted as different surface covering. The most definitive test of the theory would be direct flyby or rendezvous of a fast-rotating body, as this could obtain high resolution imaging that could specifically map out the surface morphology of the body as a function of latitude.

\section{Conclusions}

This study uses a spherical model to represent a regolith covered small asteroid.  Coulomb theory as well as DEM simulations are used to understand the failure mechanism of the regolith as the spin rate increases.  It is observed that regardless of the cohesive strength of the regolith shell, even when the body has reached its critical spin rate, the poles are always covered with regolith.  The theory is developed in a non-dimensional way, and different failure conditions and locations on the surface are found as a function of these parameters and at different spin rates. For low cohesion (signified by a low bond number), it is found that failure should occur at mid-latitudes via mass wasting. At high bond numbers failure occurs at the equator through tensile breakage of cohesive bonds. At bond numbers between these limits, a region of more complex failure is found where both mass wasting and fission can occur. Spectroscopic and thermal observations of small fast rotators can constrain the extent of regolith coverage on their surfaces.

\section{Acknowledgements}
P.S. would like to thank Audrey Thirouin for the animated discussion about her findings at Lowell Observatory that prompted the question about regolith on small super-fast rotators.  
P.S. and D.J.S. would like to acknowledge support from NASA grant 80NSSC18K0491 and support from NASA's SSERVI program. The comments of an anonymous referee are greatly appreciated and have improved the quality of this paper. 

\section*{References}
\bibliographystyle{/Users/paul/Documents/UCB/Manuscripts/elsarticle-template/model2-names}
\bibliography{/Users/paul/Documents/UCB/Meeting/psbib}

\end{document}